\documentclass[conference]{IEEEtran}
\IEEEoverridecommandlockouts

\usepackage{cite}
\usepackage{amsmath,amssymb,amsfonts}
\usepackage{algorithmic}
\usepackage{graphicx}
\usepackage{textcomp}
\usepackage{xcolor}
\usepackage{url}
\usepackage[nobiblatex]{xurl}
\usepackage{multirow}
\usepackage{longtable}
\usepackage{tabularx} 
\usepackage{booktabs}
\usepackage{array}
\usepackage{placeins}
\usepackage{enumitem}
\usepackage{array}
\usepackage{subfigure}
\usepackage{caption} 
\renewcommand{\baselinestretch}{0.98}
\usepackage{amsmath}
\usepackage{graphicx}
\usepackage{booktabs}

\def\BibTeX{{\rm B\kern-.05em{\sc i\kern-.025em b}\kern-.08em
    T\kern-.1667em\lower.7ex\hbox{E}\kern-.125emX}}
\begin{document}

\title{Towards Flexible Spectrum Access: Data-Driven Insights into Spectrum Demand\\}

\author{\IEEEauthorblockN{Mohamad Alkadamani\IEEEauthorrefmark{1}\IEEEauthorrefmark{2}\IEEEauthorrefmark{3}, Amir Ghasemi\IEEEauthorrefmark{2}, and Halim Yanikomeroglu\IEEEauthorrefmark{3}}
\IEEEauthorblockA{\IEEEauthorrefmark{2}Communications Research Centre, Ottawa, Ontario, Canada}
\IEEEauthorblockA{\IEEEauthorrefmark{3}Carleton University, Ottawa, Ontario, Canada}
\IEEEauthorblockA{\IEEEauthorrefmark{1}Corresponding Email: mohamad.alkadamani@ised-isde.gc.ca}
}

\maketitle

\begin{abstract}In the diverse landscape of 6G networks, where wireless connectivity demands surge and spectrum resources remain limited, flexible spectrum access becomes paramount. The success of crafting such schemes hinges on our ability to accurately characterize spectrum demand patterns across space and time. This paper presents a data-driven methodology for estimating spectrum demand variations over space and identifying key drivers of these variations in the mobile broadband landscape. By leveraging geospatial analytics and machine learning, the methodology is applied to a case study in Canada to estimate spectrum demand dynamics in urban regions. Our proposed model captures 70\% of the variability in spectrum demand when trained on one urban area and tested on another. These insights empower regulators to navigate the complexities of 6G networks and devise effective policies to meet future network demands.
\end{abstract}

\section{Introduction}
\label{Introduction}
As we approach the 6G era, wireless connectivity demand is expected to surge, increasing the need for spectrum \cite{itu2015}. Simultaneously, diverse 6G applications—like the Internet of Things (IoT), augmented reality (AR), and smart city infrastructure—will create more varied and complex spectrum demands, with some areas requiring high bandwidth and others needing low-latency connections \cite{ofcom2021_spectrum_strategy, itu2023_gsr}. This will make the variation in spectrum demand more apparent, necessitating flexible and adaptive spectrum management to meet both the growing demand and the heterogeneous usage patterns.

As these varied demands become more pronounced, the limitations of current spectrum allocation methods, which often involve assigning fixed, large service areas with long-term licenses, become increasingly evident. The effectiveness of any new approach hinges on accurately characterizing spectrum demand across both spatial and temporal dimensions. This task is complicated by the fact that network traffic data, crucial for assessing spectrum demand, is typically controlled by mobile network operators (MNOs) and is not made publicly available. As a result, regulators often rely on estimation models to inform their decisions.

Existing models for estimating spectrum demand often rely on broad theoretical assumptions and market studies. While these models provide a general overview, they fail to capture the intricate variations in spectrum demand that occur within smaller, localized areas. This lack of granularity leads to a one-size-fits-all approach, which overlooks the unique characteristics and demand patterns of specific areas. Moreover, there has been limited research focused on harnessing available real-world data to understand the fine-grained variations in spectrum demand and to estimate spectrum needs at the local level.

To address these limitations, this research introduces a comprehensive data-driven methodology that integrates geospatial analytics and machine learning to accurately estimate spectrum demand at localized levels. Central to our methodology is the development and validation of a proxy for spectrum demand, which is derived from real MNO traffic data. In this context, a proxy is an indicator that represents spectrum demand without directly relying on proprietary MNO traffic data. The methodology is validated through testing in two urban regions in Canada, demonstrating its robustness and effectiveness in real-world scenarios. By providing a more precise understanding of localized spectrum demand, this approach equips regulators with the insights needed to develop flexible and adaptive spectrum allocation strategies, which are essential for meeting the diverse and growing demands of 6G networks.

The remainder of this paper is structured as follows: Section~\ref{Related Work} provides a review of related works and identifies existing gaps. In Section~\ref{Methodology}, we outline our methodology, highlighting Proxy Development and Validation, Feature Engineering, and Spectrum Demand Estimation Modeling. Section~\ref{Proxy Validation} elaborates on the steps involved in proxy development and validation. Section~\ref{Problem Formulation} describes the problem formulation. Section~\ref{Feature Engineering} describes the datasets used in this study, as well as the steps required for data processing and feature engineering. Section~\ref{Spectrum Demand Estimation Modeling} elaborates on the modeling process, including performance evaluation and results discussion. Finally, Section~\ref{conclusion} summarizes the main contributions of the study.

\section{Related Work}\label{Related Work}
Existing research has predominantly focused on macro-level spectrum demand estimation, often relying on broad theoretical assumptions such as technological spectral efficiency and basic demographic indicators like population density or projected user growth. These models typically forecast demand over extended periods, usually 5 to 10 years. For example, the GSMA's global report \cite{gsma2021_mid_band} estimates mid-band spectrum requirements for 36 cities worldwide from 2025 to 2030. These estimates are based on a spectrum demand model that incorporates assumptions about user data rates, varying population densities, peak-time concurrent demand, and the proportion of traffic offloaded to high bands. Similarly, the ITU-R methodology \cite{itu2013_m1768} offers a comprehensive approach for calculating spectrum requirements for International Mobile Telecommunications (IMT), analyzing market data, service usage patterns, and traffic distribution across various radio environments and service categories to determine the necessary spectrum for projected IMT services. The FCC's empirical analysis \cite{fcc2010_mobile_broadband} estimates spectrum demands by analyzing industry forecasts of mobile data growth, adjusting for network capacity improvements through cell site expansion and spectral efficiency enhancements. Additionally, Wibisono and Elian \cite{wibisono2015} integrate statistical and trend-based inputs from the ITU and FCC models, combining site utilization factors and capacity calculations to estimate spectrum demand for mobile broadband. In another study, Irnich and Walke \cite{irnich2004spectrum} employ analytical queuing models and traffic behavior observations to estimate spectrum requirements for wireless systems. A more recent methodology detailed by Jaramillo et al. \cite{jaramillo2021spectrum} combines market studies with a forecasting model to predict spectrum needs, using key IoT drivers through a five-step process that includes traffic load calculation, capacity determination for QoS, and spectrum estimation adjusted for spectral efficiency. While these methodologies provide valuable insights, they fall short in capturing the granular, context-specific nature of spectrum demand, which is increasingly crucial for adapting to the localized requirements of next-generation networks. 

In contrast, the studies in \cite{parekh2023_data_driven} and \cite{parekh2023_data_driven2} take a data-driven approach to estimating spectrum demand at local levels. These works utilize a variety of data types as features in machine learning models to predict local spectrum demand. However, they rely on proxies that are selected based on assumptions rather than through a rigorous, data-driven process. This reliance introduces potential biases and limits the overall robustness of their findings. Unlike these previous studies, our research develops proxies directly from real traffic data and integrates detailed deployment data for precise spatial mapping. This comprehensive approach enhances the accuracy, reliability, and robustness of spectrum demand estimation, making it better suited to address the specific requirements of next-generation networks.

\begin{figure*}
\includegraphics[width=\linewidth]{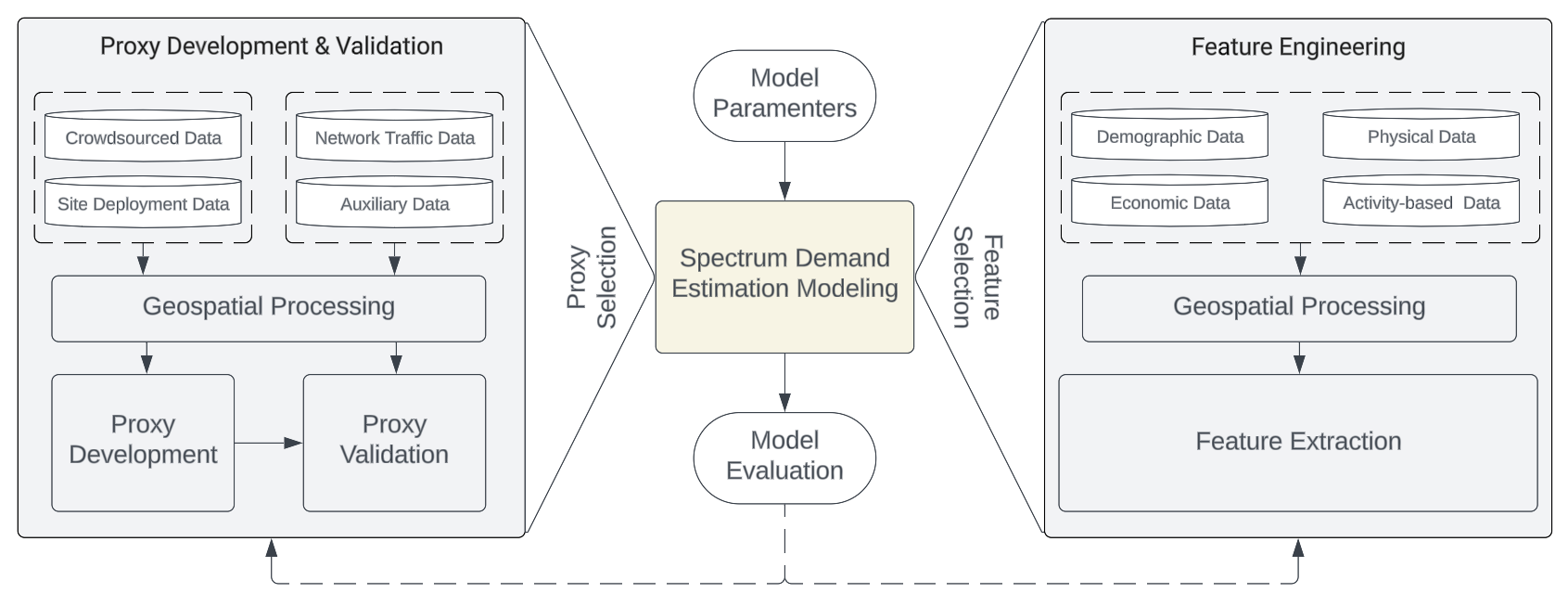}
\caption{Overview of the spectrum demand estimation methodology.}
\label{fig:ADDLframework}
\end{figure*}

\section{Methodology}\label{Methodology}

The data-driven methodology to estimate spectrum demand is organized into three critical components: Proxy Development and Validation, Feature Engineering, and Spectrum Demand Estimation Modeling. A schematic overview of these components is shown in Figure \ref{fig:ADDLframework}, clearly delineating the process flow and the integration of each part into the overall methodology.

\subsection{Proxy Development and Validation}
In this component, a proxy that represents spectrum demand is developed. This proxy is created using data that is typically publicly accessible and is then rigorously validated against proprietary MNO traffic data. The validation process ensures that the proxy accurately reflects real-world network traffic, providing a reliable representation of spectrum demand. Four primary data categories are utilized in the development and validation of this proxy:

\begin{itemize}
\item \textbf{Site deployment data:} includes technical details of mobile network sites, such as base station locations, transmitted power, and center frequencies. Proxies derived from this data represent network capacity and indirectly reflect spectrum demand.
\item \textbf{Crowdsourced measurement data:} offers end-user performance metrics like download/upload speeds and latency, collected through crowdsourcing platforms (e.g., Tutela, OpenSignal). These proxies link spectrum usage to user experiences, providing insights into Quality of Service (QoS).
\item \textbf{Traffic data:} directly sourced from operating networks, this data reveals network load and traffic patterns, serving as the ground truth for proxy validation.
\item \textbf{Auxiliary data:} includes supplementary information, such as demographic data, to refine proxy accuracy.
\end{itemize}

To facilitate analysis, these datasets are geospatially processed to ensure they are represented at a consistent spatial level. Some datasets, such as site deployment data, require the use of propagation models to achieve spatial representation. Others, like crowdsourced measurement data, already have spatial representation but need to be aligned to the appropriate geographical unit. This alignment allows for the accurate comparison of indicators across the datasets.

Proxy validation is then conducted, typically through correlation analysis, to identify which of these indicators have a high correlation with the MNO traffic data, which represents the actual spectrum demand as observed in the network. The validated proxy ultimately becomes the primary target variable for spectrum demand estimation modeling.

\subsection{Feature Engineering}
Features are the variables believed to possess predictive power for spectrum demand, and they are sourced from various datasets. The feature engineering process involves selecting, processing, and aligning these features to a consistent geographical level for effective spectrum demand modeling. Four primary data types are employed to develop features:

\begin{itemize}
    \item Demographic data: includes insights such as age distribution, population density, household composition, and employment density.
    \item Economic data: covers information like income levels, employment rates, and business establishments.
    \item Physical data: provides details on infrastructure (roads, public transport), terrain, and building footprints.
    \item Activity-based data: reflects population dynamics, such as commuting patterns and commercial activities.
\end{itemize}

These data types vary in geographical granularity, necessitating geospatial processing to standardize them to a consistent level. This involves disaggregating data from larger units or aggregating it from smaller ones. Once aligned, the features are evaluated for their predictive power through correlation analysis and feature importance ranking, selecting those with the highest impact for spectrum demand modeling.

\subsection{Spectrum Demand Estimation Modeling}
The selected proxy and features are integrated into a machine learning model through a structured process. This begins with data preparation, where the data is cleaned and formatted for modeling. Next, an appropriate model is selected, followed by the design of a training and testing strategy. The model is then fine-tuned, trained, and tested. During this phase, initial predictions are used to iteratively refine the feature engineering and proxy development processes. Finally, feature importance analysis is conducted to identify the most influential factors, ensuring the model is both accurate and explainable in estimating spectrum demand.

\section{Proxy Validation}\label{Proxy Validation}

\subsection{Spectrum Demand Indicator}
This study uses mobile traffic data from 2,799 LTE cells in Ottawa, Ontario, provided by a leading Canadian MNO, as the validation benchmark. The dataset, spanning seven primary mid-band spectrum bands, includes hourly aggregated download throughput values to calculate the spectrum demand indicator. LTE data was chosen because it is readily available and reliable in reflecting actual demand. While 5G data is still emerging and 6G data is not yet available, focusing on non-technology-specific drivers ensures that the data-driven approach remains adaptable and relevant as new technologies emerge.

For spatial analysis, the area covered by the network deployment was divided into 1.5 x 1.5 km grid cells, which served as the primary geographical unit for validation. The demand indicator was spatially represented by mapping download throughput data from individual cells to these grid cells. Coverage areas for each cell were first estimated using the extended-Hata propagation model, based on transmission parameters. The throughput values from all overlapping cell coverage areas were then aggregated and assigned to the corresponding grid cells, providing a spatial representation of spectrum demand across the study area. To further enhance spatial accuracy, the aggregated throughput was weighted using crowdsourced LTE user data, which reflects the concentration of data demand across different areas. This approach ensures a detailed and accurate depiction of where spectrum demand is most concentrated, capturing the nuances of real-world network usage patterns.

\subsection{Spectrum Demand Proxy}
Among various potential proxies considered for the spectrum demand indicator, the focus was placed on one derived from deployment data due to its strong correlation with the spectrum demand indicator and its wide availability from public sources \cite{ISED_spectrum_data}. The selected proxy, Total Deployed Bandwidth, reflects the density and variation of spectrum deployed by MNOs. The derivation process closely follows that of the spectrum demand indicator. Coverage for each tower was estimated using the extended-Hata propagation model, based on transmission data from deployment information. The deployed bandwidth for each tower was then mapped to its corresponding coverage area. To calculate the total deployed bandwidth per grid cell, bandwidth values from all overlapping tower coverage areas were aggregated.

To enhance spatial accuracy, a similar weighting method to that used for the spectrum demand indicator was applied. High-resolution satellite nighttime light (NTL) images from NASA were utilized to assess economic activity in each grid cell \cite{eog_viirs_nightlights}. The median light luminosity value was extracted, normalized, and multiplied by the deployed bandwidth to reflect the intensity of economic activity. This weighted deployed bandwidth per grid cell served as the final spectrum demand proxy, offering a detailed and accurate view of where spectrum demand is most concentrated.

\subsection{Proxy Validation}

For proxy validation, an Ordinary Least Squares (OLS) regression analysis was conducted to assess the linear correlation between the selected proxy and the spectrum demand indicator. The R-squared value for the regression model, which represents the proportion of variance in the dependent variable (spectrum demand indicator) that can be predicted from the independent variable (proxy), was calculated as 0.763. This result indicates that 76.3\% of the variability in download throughput can be explained by changes in the weighted deployed bandwidth. The relationship between the proxy and the demand indicator is illustrated in Figure \ref{fig}. The scatter plot shows a positive alignment between the proxy and the demand indicator, suggesting a strong correlation between the two.

\begin{figure}[h]
\includegraphics[width=8.5cm]{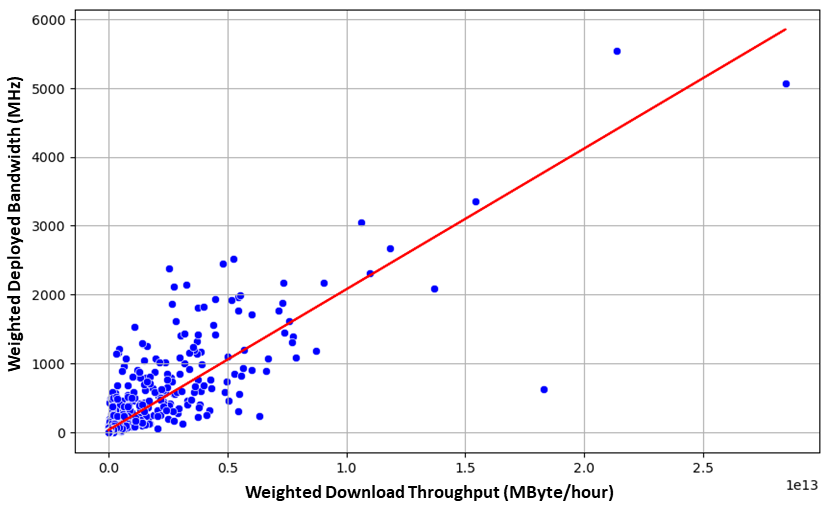}
\caption{Scatter plot comparing deployed bandwidth to download throughput.}
\label{fig:proxyvald}
\end{figure}

\section{Problem Formulation}\label{Problem Formulation}
To quantify variations in spectrum demand, the problem is framed as a regression analysis, using non-technical attributes as input features and the validated proxy as the target variable. Both the target variable and input features are spatially mapped by overlaying a grid cell map onto the areas of interest, with each grid cell measuring 1.5 x 1.5 km. This grid size provides high granularity to capture demand variation, while also allowing for aggregation to lower granularity if needed.

The study focuses on urban regions, which typically exhibit higher spectrum demand, particularly in core urban areas. However, the outskirts are also included to highlight the variation between core and peripheral areas. The analysis covers two major urban regions in Canada: the Greater Toronto Area (GTA) and Vancouver City, selected for their diverse urban densities and distinct geographic and demographic characteristics.

\begin{figure}[h]
\includegraphics[width=8.5cm]{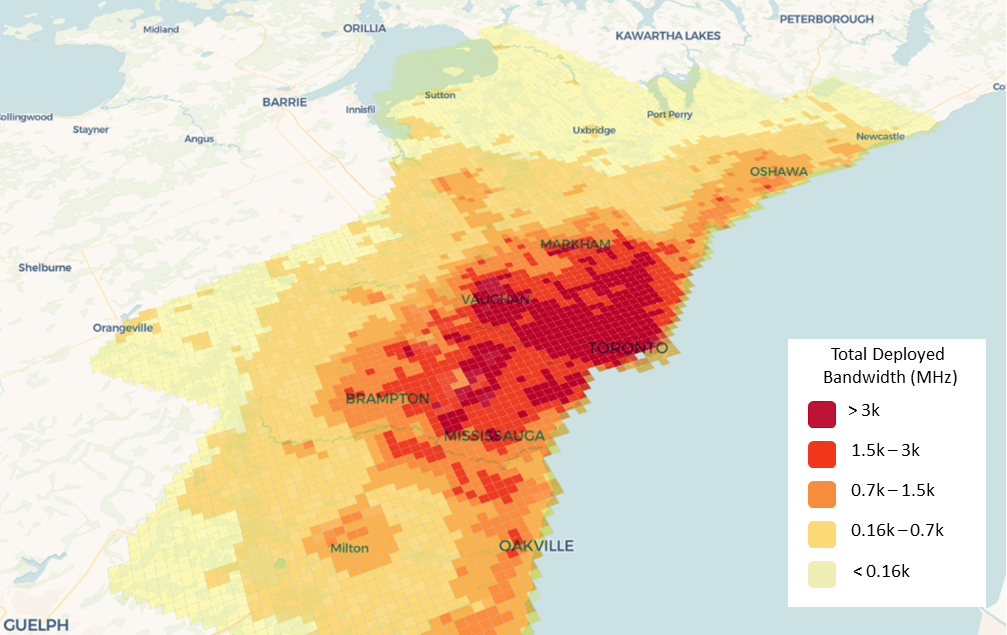}
\caption{Heatmap of total deployed bandwidth in the GTA.}
\label{fig}
\end{figure}

Figure \ref{fig} presents a sample grid overlaid on the GTA, along with a heatmap of total deployed bandwidth. The heatmap illustrates that bandwidth deployment is most concentrated in the core of the GTA and key surrounding communities, while peripheral areas show significantly lower deployment intensity. This spatial variation underscores the uneven distribution of spectrum demand across the region, aligning with the study's objective of examining local variations in spectrum demand.

\section{Feature Engineering}\label{Feature Engineering}

To develop effective machine learning models, a rigorous feature selection process was employed to identify the most predictive input variables. An ensemble approach was used, combining several feature ranking techniques: Random Forest Regressor, Gradient Boosting Regressor, Lasso Regression, Ridge Regression, and Feature Permutation Importance. Each method independently assessed feature importance, leveraging different data aspects to highlight variables with the greatest impact on prediction accuracy. The importance scores from each method were normalized and averaged to create an aggregate importance score for each feature. This ensemble approach provided a comprehensive and balanced evaluation of feature importance, enhancing the model's robustness and interpretability.

\subsection{Demographic Data}
The primary source of demographic data is the 2021 Census from Statistics Canada. This data encompasses various geographical areas, including census subdivisions (CSDs) and dissemination areas (DAs), which are the specific geographic units for which census data are disseminated. The dataset includes:

\begin{itemize}[label=-]
    \item Population density: measures the number of people per unit area at their residence (nighttime population).
    \item Socioeconomic indicators: includes levels of income, education, and age group distributions. These factors are often correlated with technological usage and can influence spectrum demand.
    \item Industry presence: utilizes data from the North American Industry Classification System (NAICS) \cite{naics2017} to gauge industrial density. All industries within a grid cell are aggregated to produce a single value representing industry density.
\end{itemize}

\subsection{Daytime Population Data}

This dataset estimates the population present in a geographic area during daytime hours, distinguishing between those at home and those at work. Unlike the nighttime population data from Statistics Canada, which reflects where people reside, this dataset captures the dynamic distribution of the population during work hours.

\subsection{Physical and Environmental Data}

\begin{itemize}[label=-]
\item \textbf{Building Density:}
this includes data from Microsoft Building Footprints \cite{microsoft_building_footprints}, which identifies buildings using AI-assisted analysis of satellite imagery, and data from OpenStreetMap on non-residential buildings, which provides a broader view of non-residential infrastructure density. For both datasets, key metrics such as building density, the total number of buildings, and the total area covered by buildings within each geographic unit were calculated and incorporated as multiple features in the dataset.
\item \textbf{Infrastructure Data:} 
this includes metrics such as the total length of streets, the number of transportation hubs, and points of interest within each grid cell, which serve as indicators of population concentration and activity levels. These metrics were also sourced from OpenStreetMap \cite{statscan_road_network}, \cite{openstreetmap_poi}.

\end{itemize}

\subsection{Commute Data}
The average distance traveled per day by different family groups, providing insights into mobility patterns that can affect local spectrum demand.

A comprehensive list of data types and their respective sources can be seen in Table \ref{tab:feature_engineering_data}.

\begin{table}[h]
\centering
\caption{Feature Engineering Data Summary}
\label{tab:feature_engineering_data}
\begin{tabular}{@{}p{0.25\columnwidth}p{0.65\columnwidth}@{}}
\toprule
\textbf{Data Type} & \textbf{Description} \\
\midrule
Population Density (Statistics Canada) & Count of individuals per given area during nighttime\\
\midrule
Socio-economic Indicators (Statistics Canada) & Data reflecting various socio-economic metrics such as income levels, education attainment, and age demographics \\
\midrule
Industry Presence (Statistics Canada) & Density of different industrial sectors, as defined by NACIS, within a specific geographic area \\
\midrule
Daytime Population (Environics Analytics) & Count of individuals per given area during daytime \\
\midrule
Building Footprints (Microsoft) & Geospatial data representing the outlines of building structures \\
\midrule
Non-Residential Buildings (OpenStreetMap) & Geospatial data representing the footprints of non-residential structures \\
\midrule
Environment (OpenStreetMap) & Geospatial data capturing road networks, distribution of transportation hubs, and points of interest (POI)  \\
\midrule
Commute Distance (Statistics Canada) & Data providing insights into commuting distances for various age groups \\
\bottomrule
\end{tabular}
\end{table}

\section{Spectrum Demand Estimation Modeling}\label{Spectrum Demand Estimation Modeling}

This section examines two methodologies for predicting spectrum demand: a simple linear model and more advanced machine learning (ML) models. The effectiveness of these approaches is evaluated by assessing their ability to estimate spectrum demand based on various urban metrics.

\subsection{Linear Model as Baseline}
The linear model is used as the baseline in this study, relying on a single feature that demonstrates the strongest predictive power among all input features. Specifically, the number of transportation hubs within each grid cell shows the highest correlation with the weighted deployed bandwidth. This baseline model was developed and evaluated for a combined urban regions scenario, which includes data from both the GTA and Vancouver.

\begin{figure}[h]
\includegraphics[width=8.5cm]{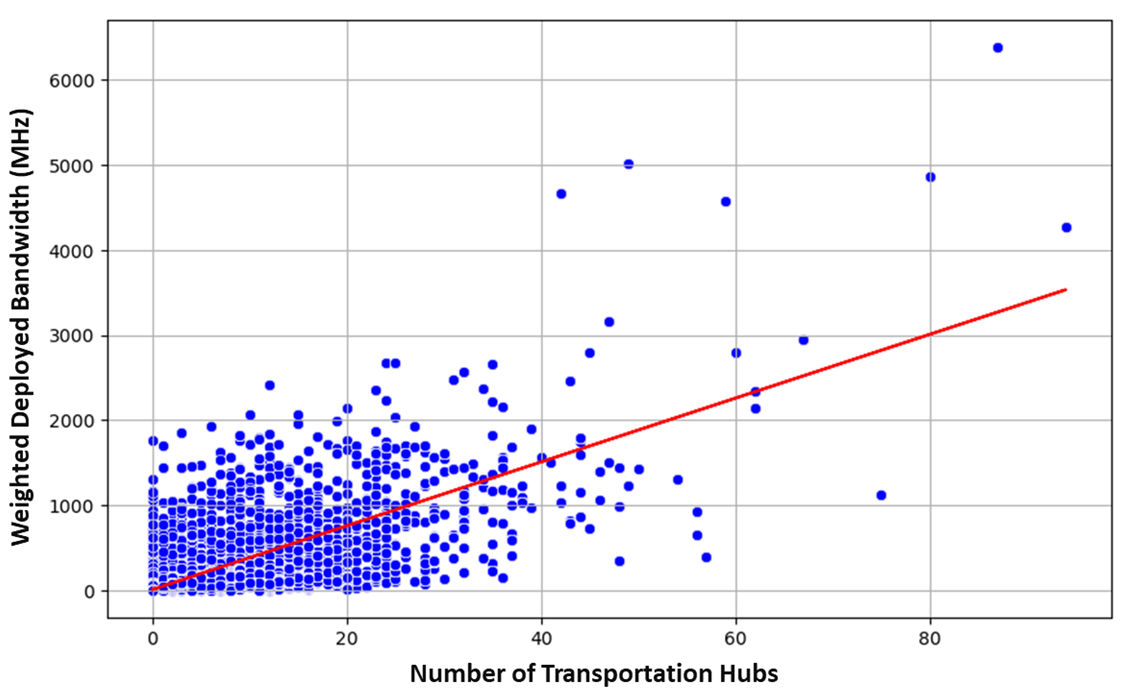}
\caption{Scatter plot comparing the number of transportation hubs and weighted deployed bandwidth.}
\label{fig:transport}
\end{figure}

Figure \ref{fig:transport} illustrates a scatter plot depicting the relationship between the number of transportation hubs and the weighted deployed bandwidth. The analysis reveals a positive correlation with an
R-squared value of 0.58, indicating that 58\% of the variance in the total weighted deployed bandwidth can be explained by the number of transportation hubs. This correlation underscores the relevance of transportation hubs as a predictor, particularly in urban settings.

\subsection{Machine Learning Models}

To ensure the accuracy of our machine learning models, particularly for geospatial data, we employed K-means clustering using the latitude and longitude coordinates of each grid cell. This step was crucial as geospatial datasets often exhibit spatial dependencies, meaning that selecting data points from one area could introduce bias into the results. By utilizing K-means clustering, we effectively grouped similar geographical regions together, creating 15 distinct clusters. Subsequently, we randomly divided the data within each cluster into two subsets: 80\% for training and 20\% for testing. This approach ensured that the geographical diversity within the dataset was well-represented, allowing for a more robust training and testing dataset. 

Two ML models were employed, Ridge Regression and GBR, to estimate spectrum demand within urban environments. Ridge Regression was selected for its ability to manage multicollinearity through regularization. The model uses a regularization parameter, alpha, set at 0.1, which helps prevent overfitting by penalizing large coefficients. This is particularly useful in urban data sets where predictors are often correlated, ensuring the model remains robust and generalizable. GBR, on the other hand, was chosen for its high efficiency and predictive accuracy. Both models were evaluated using 5-fold cross-validation to robustly estimate their performance. The model builds an ensemble of weak prediction models, typically decision trees, in a stage-wise fashion. It optimizes for a loss function, in this case, the mean squared error, which is particularly effective for continuous output variables like spectrum demand. Both models were evaluated using 5-fold cross-validation to robustly estimate their performance. The results from these models are compared to the baseline linear model to evaluate improvements in predictive accuracy and to identify the most influential predictors of spectrum demand. 

\subsection{Performance Evaluation} 

The evaluation of ML models involves assessing their performance using key metrics such as R-squared and Root Mean Square Error (RMSE). RMSE quantifies the average deviation of predicted values from actual values, with lower RMSE values indicating better model performance.

The performance of the ML models was examined in predicting the weighted deployed bandwidth of a major operator in Canada across two urban regions. This is the same operator used in the proxy validation process to ensure consistency throughout the analysis. Two distinct scenarios were explored. The first scenario involves training and testing the models on combined urban regions, encompassing data from both GTA and Vancouver. This approach provides a comprehensive view of spectrum demand patterns across diverse urban landscapes. In contrast, the second scenario entails training the models on data exclusively from GTA and testing them on Vancouver. This scenario examines the models' ability to generalize across urban regions with different characteristics.

Table \ref{tab:model_performance} presents the performance metrics for the two scenarios for both ML models, the Ridge Regression and GBR models.

\begin{table}[h]
\centering
\caption{Summary of ML Models' Performance}
\label{tab:model_performance}
\begin{tabular}{@{}ccc@{}}
\toprule
\textbf{Model} & \textbf{Scenario} & \textbf{Performance} \\ \midrule
Ridge Regression & Combined Urban Regions & $R^2 = 0.64$, RMSE = 0.96 \\
Ridge Regression & GTA - Vancouver & $R^2 = 0.53$, RMSE = 1.9 \\
GBR & Combined Urban Regions & $R^2 = 0.81$, RMSE = 0.51 \\
GBR & GTA - Vancouver & $R^2 = 0.70$, RMSE = 0.93 \\ \bottomrule
\end{tabular}
\end{table}

The models employed in this study are well-established and widely used in predictive modeling, making them both practical and accessible for implementation. These models are computationally efficient, even with large datasets, and can be easily integrated into existing data analysis pipelines. Ridge Regression, with its simplicity and ability to handle multicollinearity, requires minimal computational resources, while GBR, though slightly more resource-intensive, offers high predictive accuracy and scalability. Both models can be run on standard computational hardware and are supported by popular machine learning libraries, ensuring ease of deployment in various spectrum management systems.

\subsection{Discussion}

The performance evaluation sheds light on the predictive capabilities of the Ridge Regression and GBR models across diverse scenarios.

\subsubsection{Combined Urban Regions}

In the scenario where training and testing were conducted on combined urban regions, both machine learning models demonstrated strong performance, surpassing the baseline linear model. The GBR model, in particular, exhibited exceptional predictive accuracy, achieving an R-squared value of 0.81 and a low RMSE of 0.51. This represents a significant 23\% improvement over the baseline model, underscoring the superior capability of machine learning models to capture intricate relationships within the data.

\subsubsection{Training on GTA, Testing on Vancouver (GTA-Vancouver)}

When the models were trained on data from the GTA and tested on Vancouver, a decline in performance was observed compared to the combined urban regions scenario. Despite this, the GBR model continued to estimate spectrum demand effectively. Notably, the GBR model successfully identified common demand indicators that influence spectrum demand across different urban regions and applied them effectively. This resilience highlights the model's adaptability and its capacity to capture essential features that contribute to spectrum demand variation in diverse urban settings.

The key features influencing download bandwidth estimation were further analyzed using the gain-based feature importance algorithm \cite{Manish2016}. This analysis assessed the relative impact of different input features on the overall loss function. Figure \ref{fig} illustrates the twelve most influential features identified by the GBR model for both scenarios: (a) Combined Urban Regions and (b) GTA-Vancouver.

In the combined scenario, most features exhibited a similar level of importance, indicating their collective contribution to understanding the variability of deployed bandwidth. In contrast, in the GTA-Vancouver scenario, the importance of features significantly declined after the first six, suggesting that these initial features were either predominantly utilized or shared common characteristics between the two regions. To further explore the impact of these key features, additional tests were conducted using a reduced set of features.

First, the model's performance was evaluated using only the five most important features identified in the GTA-Vancouver scenario. The results indicated that the model maintained a relatively high level of accuracy, with the R-squared value decreasing slightly from 0.70 to 0.68. This suggests that these top five features capture most of the variability in spectrum demand, enabling the model to remain both interpretable and streamlined without a substantial loss in predictive power. The model was then tested using only the single most important feature, the number of roads. In this case, the R-squared value further decreased to 0.61. While this reduction in accuracy was expected, the model still provided valuable insights, highlighting the significant influence of this feature on spectrum demand across the two regions. These findings underscore the importance of feature selection in balancing model simplicity and performance. By focusing on a smaller set of key features, the model becomes more interpretable and easier to apply in practical settings, particularly for decision-makers who require straightforward explanations of the factors driving spectrum demand.

An additional observation is that, despite the widespread use of nighttime population density as an indicator in many models, the data in this study shows it performs poorly in predicting spectrum demand. In contrast, daytime population density performed well in both scenarios, highlighting that spectrum demand is more closely tied to daytime activity patterns. This finding emphasizes the importance of considering where people are active during the day rather than relying solely on nighttime population data.
\begin{figure}[h]
    \centering
    \subfigure[Combined urban regions scenario.]{
        \includegraphics[width=8.5cm]{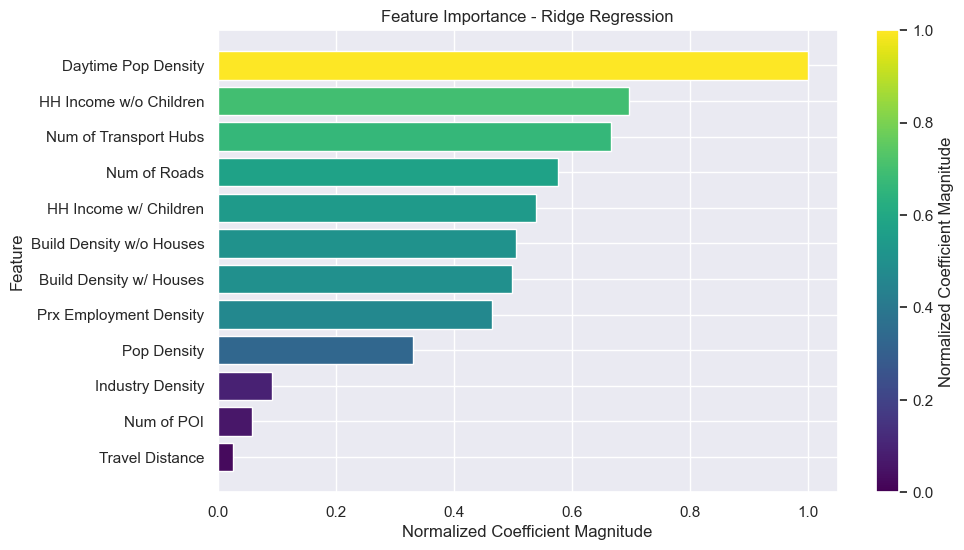}
        \label{fig:proxy1}
    }
    \subfigure[GTA-Vancouver scenario.]{
        \includegraphics[width=8.5cm]{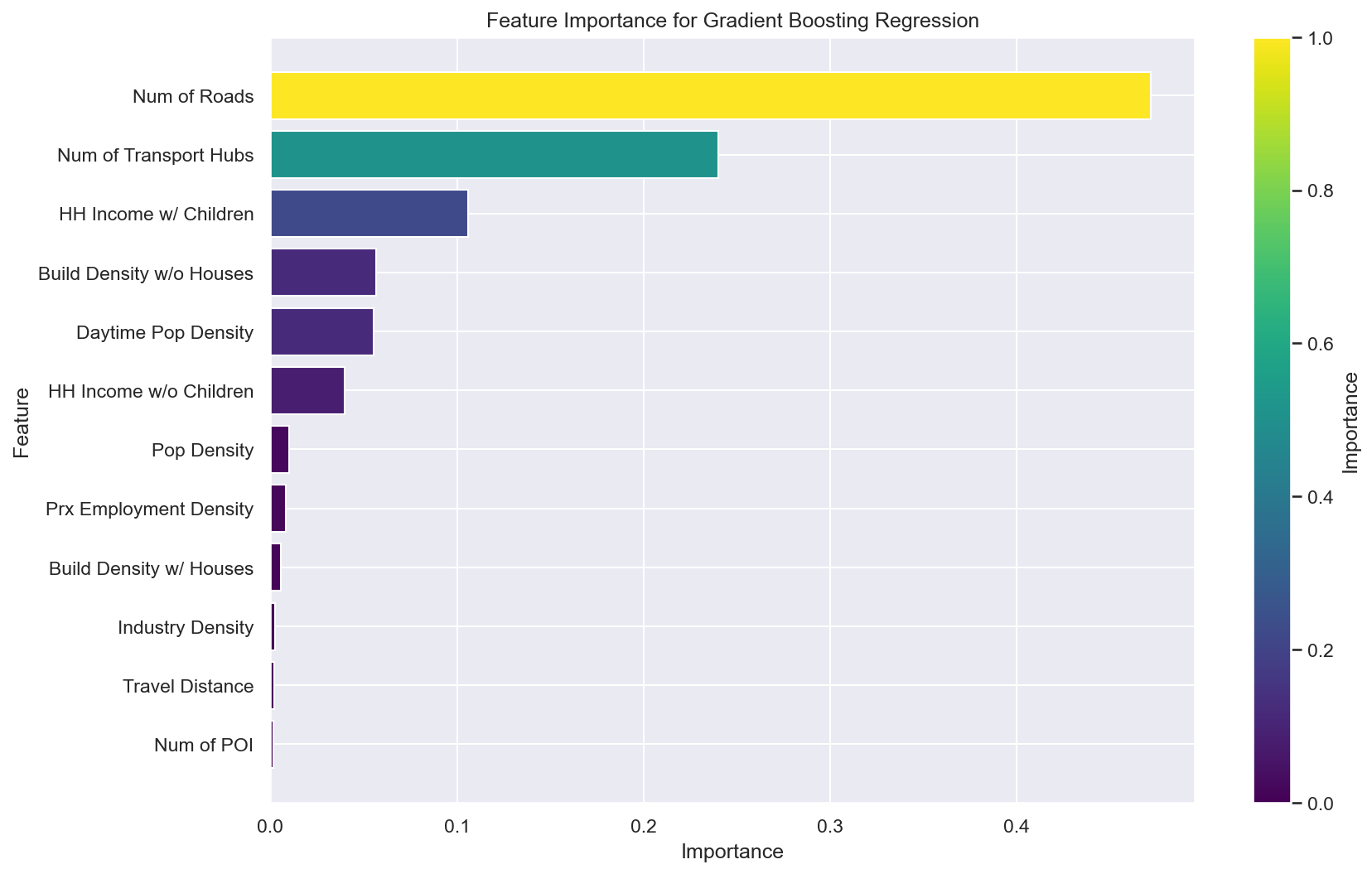}
        \label{fig:proxy2}
    }
    \caption{Feature importance for spectrum demand estimation.}
    \label{fig:feature}
\end{figure}

\section{Conclusion}\label{conclusion}

This study presents a comprehensive, data-driven methodology that integrates machine learning and geospatial analytics to precisely estimate spectrum demand and identify key drivers within the mobile broadband landscape. By leveraging real network traffic data, spectrum demand proxies were developed, and features were engineered based on demographic, social, and physical characteristics. The case studies conducted in Canada demonstrated the robustness of this approach, achieving an R-squared value of 0.70 when trained on one urban region and tested on another, and an even higher R-squared value of 0.81 when applied across combined urban regions.

These results underscore the methodology's ability to move beyond traditional, macro-level spectrum demand estimation, offering a more granular and accurate tool for policymakers. By capturing detailed spatial variations in spectrum demand, the model significantly enhances the understanding of key demand drivers, enabling the development of tailored and responsive spectrum policies. This approach not only addresses current spectrum management challenges but also lays a strong foundation for future policy decisions in the evolving landscape of mobile broadband.

\bibliographystyle{IEEEtran}
\bibliography{biblio}

@techreport{itu2015,
    author = {{International Telecommunication Union}},
    title = {{IMT Traffic Estimates
for the Years 2020 to 2030}},
    year = {2015},
    institution = {ITU-R},
    note = {\url{https://www.itu.int/dms_pub/itu-r/opb/rep/R-REP-M.2370-2015-PDF-E.pdf}}
}

@techreport{ofcom2021_spectrum_strategy,
    author = {Ofcom},
    title = {{Supporting the UK’s Wireless Future: Our Spectrum Management Strategy for the 2020s}},
    year = {2021},
    institution = {Ofcom},
    note = {\url{https://www.ofcom.org.uk/__data/assets/pdf_file/0017/222173/spectrum-strategy-statement.pdf}}
}

@techreport{itu2023_gsr,
    author = {{International Telecommunication Union}},
    title = {{GSR-23: Chairman's Report - Global Symposium for Regulators}},
    year = {2023},
    note = {\url{https://www.itu.int/itu-d/meetings/gsr-23/wp-content/uploads/sites/20/2023/07/GSR23_Chairman-Report_Final_English.pdf}}
}

@techreport{gsma2021_mid_band,
    author = {{GSMA and Coleago Consulting}},
    title = {{Estimating Mid-band Spectrum Needs in the 2025-2030 Time Frame: Global Outlook}},
    year = {2021},
    institution = {GSMA},
    note = {\url{https://www.gsma.com/connectivity-for-good/spectrum/wp-content/uploads/2021/07/Estimating-Mid-Band-Spectrum-Needs.pdf}}
}

@inproceedings{wibisono2015,
    author = {G. Wibisono and B. Elian},
    title = {{Design of Spectrum Estimation Model for Mobile Broadband in Indonesia from 2015 to 2025}},
    booktitle = {IEEE European Modelling Symposium (EMS)},
    location = {Madrid, Spain},
    pages = {43-48},
    year = {2015},
    doi = {10.1109/EMS.2015.17}
}

@techreport{itu2013_m1768,
    author = {{International Telecommunication Union}},
    title = {{R-REC-M.1768-1: Methodology for Calculation of Spectrum Requirements for the Terrestrial Component of IMT}},
    year = {2013},
    institution = {ITU-R},
    note = {\url{https://www.itu.int/dms_pubrec/itu-r/rec/m/R-REC-M.1768-1-201304-I!!PDF-E.pdf}}
}

@techreport{fcc2010_mobile_broadband,
    author = {{Federal Communications Commission}},
    title = {{Mobile Broadband: The Benefits of Additional Spectrum}},
    year = {2010},
    institution = {FCC},
    note = {\url{https://transition.fcc.gov/national-broadband-plan/mobile-broadband-paper}}
}

@inproceedings{irnich2004spectrum,
  title={Spectrum Estimation Methodology for Next Generation Wireless Systems},
  author={Irnich, T. and Walke, B.},
  booktitle={IEEE 15th International Symposium on Personal, Indoor and Mobile Radio Communications},
  pages={1957--1962},
  volume={3},
  year={2004},
  organization={IEEE},
  address={Barcelona, Spain},
  doi={10.1109/PIMRC.2004.1368340}
}

@article{jaramillo2021spectrum,
  title={{Spectrum Demand Forecasting for IoT Services}},
  author={Jaramillo-Ramirez, Daniel and P{\'e}rez Cerquera, Manuel},
  journal={Future Internet},
  volume={13},
  year={2021},
  doi={10.3390/fi13090232}
}

@inproceedings{parekh2023_data_driven,
    author = {{J. Parekh, A. Ghasemi, and H. Yanikomeroglu}},
    title = {{Data-driven Modelling of Mobile Network Demand for Efficient Spectrum Management}},
    booktitle = {IEEE 34th Annual International Symposium on Personal, Indoor and Mobile Radio Communications (PIMRC)},
    location = {Toronto, ON, Canada},
    pages = {1-6},
    year = {2023},
    doi = {10.1109/PIMRC56721.2023.10294042}
}

@inproceedings{parekh2023_data_driven2,
    author = {{J. Parekh, E. Yackoboski, A. Ghasemi and H. Yanikomeroglu}},
    title = {{Modeling Local Demand for Mobile Spectrum using Large Crowdsourced Datasets}},
    booktitle = {IEEE Future Networks World Forum (FNWF)},
    location = {Baltimore, MD, USA},
    pages = {1-5},
    year = {2023},
    doi = {10.1109/FNWF58287.2023.10520414}
}

@online{ISED_spectrum_data,
    author = {{Innovation, Science and Economic Development Canada. (2023).}},
    year = {},
    title = {{Spectrum Management System Data [Data table]}},
    url = {https://smssgs.ic.gc.ca/eic/site/sms-sgs-prod.nsf/eng/h00010.html.},
    note = {}
}

@online{eog_viirs_nightlights,
    author = {{Earth Observation Group. (2020).}},
    year = {},
    title = {{Annual VIIRS Nighttime Lights v2}},
    url = {https://eogdata.mines.edu/products/vnl/#annual_v2.},
    note = {}
}

@online{naics2017,
    author = {{Statistics Canada. (2017).}},
    year = {},
    title = {{North American Industry Classification System (NAICS) Canada 2017 Version 3.0 [Data table]}},
    url ={https://www.statcan.gc.ca/en/subjects/standard/naics/2017/v3/index},
    note = {}
}

@misc{microsoft_building_footprints,
   author = {{Microsoft}},year = {2019},
   title = {{Building Footprints}},
   howpublished = "\url{ https://github.com/microsoft/CanadianBuildingFootprints.}"  
 }

@online{statscan_road_network,
    author = {{Statistics Canada. (2021).}},
    year = {},
    title = {{Road Network}},
    url = {https://www12.statcan.gc.ca/census-recensement/2011/geo/RNF-FRR/index-eng.cfm.},
    note = {}
}

@misc{openstreetmap_poi,
   author = {{OpenStreetMap Contributors}},
   title = {{Points of Interest}},
   howpublished = "\url{ https://wiki.openstreetmap.org/wiki/points_of_interest.}",
   year = {2023},
 }

@article{Manish2016,
    author = {Gulati, Pooja and Sharma, Amita and Gupta, Manish},
    title = {{Theoretical Study of Decision Tree Algorithms to Identify Pivotal Factors for Performance Improvement: A Review}},
    journal = {International Journal of Computer Applications},
    volume = {141},
    pages = {19-25},
    year = {2016},
    doi = {10.5120/ijca2016909926}
}

\end{document}